\documentclass[aps,showpacs,superscriptaddress,twocolumn,amsmath,amssymb,pre]{revtex4-1}
\usepackage{dcolumn}
\usepackage{yfonts}
\usepackage{bm}
\usepackage{amssymb}
\usepackage{graphicx}
\usepackage{color}
\usepackage{wrapfig,subfigure}

\newcommand{\ep}{\varepsilon}
\newcommand{\ka}{\varkappa}

\def\hb0{h_{\rm b}^{(0)}}
\def\p12{p_{12}({\bf q},t)}

\begin{document}

\title {Quantum critical temperature of a modulated oscillator}

\author{ Lingzhen Guo$^{1,2}$,Vittorio Peano$^{3}$, M. Marthaler$^{1,4}$, and M. I. Dykman$^{3}$}

\affiliation{$^1$Institut f\"ur Theoretische Festk\"orperphysik, Karlsruhe Institute of Technology, 76128 Karlsruhe, Germany\\
$^2$Department of Physics, Beijing Normal University, Beijing 100875, China\\
$^3$ Department of Physics and Astronomy, Michigan State University, East Lansing, MI 48824, USA\\
$^4$ DFG-Center for Functional Nanostructures (CFN),
Karlsruhe Institute of Technology, 76128 Karlsruhe, Germany}

\date{\today}

\begin{abstract}

We show that the rate of switching between the vibrational states of a modulated nonlinear oscillator is characterized by a quantum critical temperature  $T_c\propto\hbar^2$. The rate is independent of $T$ for $T<T_c$.  Above $T_c$ there emerges a quantum crossover region where the slope of the logarithm of the distribution over the oscillator states displays a kink  and 
the switching rate has the Arrhenius form with the activation energy  independent of the modulation. The results demonstrate the limitations of the real-time instanton theory of switching in systems lacking detailed balance.

\end{abstract}

\pacs{05.40.-a, 03.65.Yz,  74.50.+r, 85.25.Cp}
  \maketitle

\section{Introduction}

Thermal equilibrium systems have time reversal symmetry, which leads to detailed balance: the rates of transitions back and forth between any two states are equal \cite{Lifshitz1981a}.  Nonequilibrium systems generally do not have detailed balance. An important exception is a nonlinear oscillator additively modulated close to its eigenfrequency $\omega_0$ \cite{Drummond1980c} or parametrically modulated at $\approx 2 \omega_0$ \cite{Kryuchkyan1996}. This system has detailed balance in the rotating wave approximation (RWA), but only for zero temperature \footnote{Detailed balance emerges also in some classical and quantum models of parametrically pumped coupled modes  \protect\cite{Woo1971,Drummond1981a,*Drummond1989,*Kinsler1991,Wolinsky1988}}. For nonzero temperature the oscillator  does not have detailed balance, cf. \cite{Dykman1979a}. 

Interestingly, the detailed-balance probability distribution of the oscillator is ``fragile". The distribution for $T\to 0$ can be exponentially strongly different from that for $T=0$ for weak damping and comparatively strong modulation, where the oscillator dynamics is semiclassical \cite{Dykman1988a,Marthaler2006}. The fragility was found in the region where the oscillator has two stable vibrational states (SVSs). As its consequence, the rate of switching from a vibrational state $W_{\rm sw}$ can be exponentially increased for $T>0$ compared to its value for $T=0$. Such increase is of significant interest, as the bistability of forced vibrations of quantum nonlinear oscillators plays an important role in their applications in quantum information, and the interstate switching is attracting much attention, cf. \cite{Peano2004,*Peano2010,Katz2007,Vijay2009,*Murch2011,Mallet2009,Bishop2010,*Ginossar2012,Wilson2010}.

In this paper we show that the $T=0$ region and the region, where the lack of detailed balance in the underdamped modulated oscillator is pronounced, are separated by a  quantum temperature $T_c\propto \hbar^2$. This temperature does not show up in the standard WKB approximation, it is not related to the conventional  WKB corrections $\propto \hbar^2$. For $T>T_c$ the exponent of the switching rate $W_{\rm sw}$ changes with increasing $T$ from the semiclassical $T=0$ to the semiclassical $T\to 0$ value. The change occurs in the temperature range that scales with $\hbar$  as $\hbar/|\log \hbar|$.

The distribution of the modulated oscillator over its quantum states is formed as a result of the coupling to a thermal reservoir. The coupling leads to oscillator relaxation. In a simple picture relaxation comes from emission of excitations of the bath, for example, photons.
It is invariably accompanied by noise, because photons are emitted at random.  For $T>0$, along with the noise from photon emission, there is noise from photon absorption. Its relative intensity for low $T$ is  $\propto \bar n$, where $\bar n = [\exp(\hbar\omega_0/k_BT)-1]^{-1}$ is the oscillator Planck number. The noise leads to fluctuations of the oscillator and ultimately to switching between the SVSs over an effective barrier in phase space \cite{Dykman1988a,Marthaler2006}.
\begin{figure}[h]
\includegraphics[width=74mm]{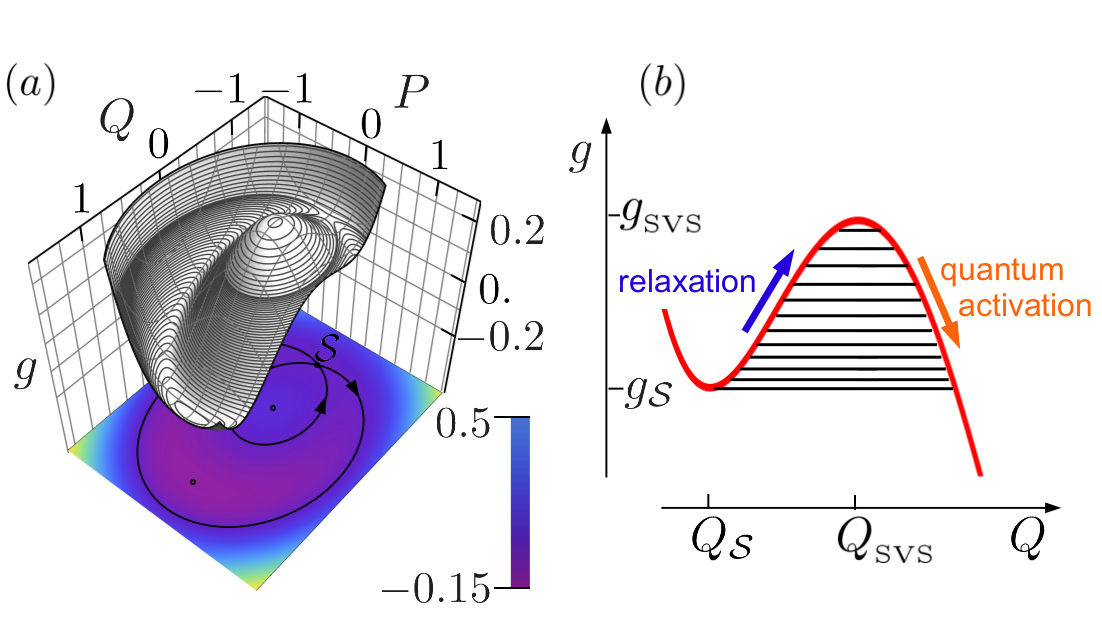}
\caption{ (a) The scaled Hamiltonian $g(Q,P)$ of the resonantly driven nonlinear oscillator; $Q$ and $P$ are
the dimensionless coordinate and momentum in the rotating frame. In the presence of weak dissipation, the local maximum and minimum of $g$ correspond to the small- and large-amplitude stable vibrational states (SVSs). The scaled field intensity is $\beta=0.02$. (b) The cross-section $g(Q,0)$; the solid lines show the quasienergy states localized about the local maximum of $g$ at $Q=Q_{_{\rm SVS}}$ for $\lambda=0.015$. The scaled quasienergies of these states $g_n$ lie between $g_{_{\rm SVS}}=g(Q_{_{\rm SVS}},0)$ and $g_{_{\cal S}}=g(Q_{_{\cal S}},0)$, where  $(Q_{_{\cal S}},P=0)$ is the saddle point of $g(Q,P)$.}
\label{fig:quasienergy}
\end{figure}

An insight into the onset of the switching rate fragility can be gained from Fig.~\ref{fig:quasienergy}, which shows the effective oscillator  Hamiltonian in the rotating frame. Its extrema correspond to the SVSs, for weak damping. Quantum states of the oscillator localized about an extremum of the Hamiltonian are analogous to intrawell states of a particle in a static potential. For weak damping, the state populations $\rho_{n}$ are determined by a balance equation. In dimensionless time $\tau$, see Eq.~(\ref{eq:master_equation}), it reads
\begin{equation}
\label{eq:balance_equation}
\dot\rho_{n} \equiv d\rho_n/d\tau=\sum_m(W_{mn}\rho_{m}-W_{nm}\rho_{n}).
\end{equation}
The dimensionless  transitions rates $W_{mn}=W_{mn}^{\rm (e)}+W_{mn}^{\rm (abs)}$ have two terms, which come from emission and absorption of bath excitations, respectively; $W_{mn}^{\rm (abs)}\propto \bar n$.

For $T=0$, the stationary probability distribution $\rho_{n}^{(0)}$ is formed by the rates $W_{mn}^{\rm (e)}$. Of interest for switching is the population of quantum states $n$ with the wave functions $\psi_n$ extending to the boundary of the classical basin of attraction to the SVS. For such states $n\gg 1$, for small $\hbar$ (we count the states off from the SVS).
In a certain parameter range the rates $W_{m\,n}^{\rm (abs)}$ of absorption-induced transitions decay with $n-m>0$ slower than $\rho^{(0)}_{n}/\rho^{(0)}_{m}$. 
Then, if we consider these transitions as a perturbation  in Eq.~(\ref{eq:balance_equation}), the correction $\sum_{m<n} W_{m\,n}^{\rm (abs)}\rho_{m}^{(0)}/\rho_n^{(0)}$ to $\dot\rho_n/\rho_n$ diverges for $n\to \infty$. This correction describes the absorption-induced flux into state $n$ from the states that are closer to the SVS. The divergence occurs even for $\bar n\to 0$, indicating the fragility of the $T=0$ solution.

The value of $T_c$ can be estimated by noticing that, for finite $\hbar$, the number $M_{}$ of  quantum states localized about an SVS is finite, $M_{}\propto \hbar^{-1}$.  
Therefore the above correction has a finite number of terms. If the characteristic ratio of the consecutive terms is $v>1$, the correction is $\propto \bar n v^{M_{}}$ for $n\sim M_{}$. It
remains small
for small $\bar n v^M$, i.e., for $T <T_c\sim  (\hbar\omega_0/k_B)/M_{}\log v\propto \hbar^{2}$.

Finding the distribution for a finite number of states requires calculating the rates $W_{nm}$ of transitions between remote states, $|n-m|\gg 1$. We do this for an oscillator modulated close to its eigenfrequency by combining the WKB and the conformal mapping techniques. 

\section{Oscillator dynamics in the rotating frame}

The Hamiltonian of the modulated oscillator reads
\begin{equation}
H_0=\frac{p^2}{2}+\frac{1}{2} \omega_0^2 q^2 + \frac{\gamma}{4} q^4 - A q \cos(\omega_F
t),  
\label{eq:Hamiltonian}
\end{equation}
where $q$ and $p$ are the oscillator coordinate and momentum. We assume that the modulation detuning $\delta\omega=\omega_F-\omega_0$ is small,  $|\delta\omega|\ll\omega_0$, as is also the nonlinearity , $|\gamma|\langle q^2\rangle\ll\omega_0^2$. The oscillator displays bistability of forced vibrations for $\gamma\,\delta\omega>0$; for concreteness we assume $\gamma,\, \delta\omega > 0$.

We switch to the rotating frame using the transformation $U(t)=\exp (-i\omega_F a^\dagger a t)$, where $a^\dagger$ and $a$ are the oscillator raising and lowering operators. We also introduce slowly varying in time dimensionless coordinate $Q$ and momentum $P$, 
$U^\dagger (t) q U(t) =C\left(Q\cos \omega_F t + P\sin \omega_F t\right)$ 
and $U^\dagger (t) p U(t)  = -C\omega_F\left( Q\sin \omega_F t - P\cos \omega_F t\right),$
with the scaling constant $C=(8\omega_F\delta\omega/3\gamma)^{1/2}$; 
\begin{equation}
\label{eq:Q_P_commutator}
[Q,P]=i\lambda, \qquad \lambda = 3\hbar \gamma/8\omega_F^2(\omega_F-\omega_0).
\end{equation}
The dimensionless parameter $\lambda$ plays the role of the Planck constant in the dynamics in the rotating frame. We assume $\lambda \ll 1$, so that this dynamics is semiclassical.

In the RWA the transformed oscillator Hamiltonian $\tilde{H}_0=[8\omega_F^2(\delta\omega)^2/3\gamma]\hat g$ is independent of time. Here, 
\begin{equation}
\label{eq:g_{e}unction}
\hat{g}=g(Q,P)= \frac{1}{4}(Q^2+P^2-1)^2 - \beta^{1/2} Q;
\end{equation}
$\beta=3|\gamma| A^2/32\omega_F^3|\omega_F-\omega_0|^3 $ is the scaled field intensity. 
Function $g(Q,P)$ is shown in Fig.~\ref{fig:quasienergy}. Where the oscillator is bistable,  $g(Q,P)$ has the form of a tilted Mexican hat with the extrema corresponding to the SVSs.

Operator $\hat{g}$ is Hermitian and has a complete set of eigenfunctions $\psi_n(Q)$, $\hat{g} \psi_n=g_n\psi_n$. The eigenvalues $g_n$ give the scaled quasienergies $\ep_n$ of the modulated oscillator  in the RWA, $\ep_n \approx  [8\omega_F^2(\delta\omega)^2/3\gamma]g_n$.  They are shown in Fig.~\ref{fig:quasienergy}(b). The spacing between the eigenvalues $g_n$ is $\propto\lambda$. The number of states between an extremum and the saddle point of $g(Q,P)$ is $M\sim \lambda^{-1} \gg 1$.

\subsection{Master equation and the transition rates}

The dynamics of the oscillator weakly coupled to a thermal reservoir can often be described in slow time by a Markov master equation for the oscillator density matrix $\rho$. To the lowest order in the coupling
\begin{eqnarray}
\label{eq:master_equation}
d\rho/d\tau= - i\lambda^{-1}[\hat{g},\rho]-\hat\kappa\rho, \qquad \tau = |\omega_F-\omega_0|t.
\end{eqnarray}
Operator $\hat\kappa\rho$ describes relaxation; the coupling-induced renormalization of the oscillator parameters is incorporated into $\omega_0,\gamma$. If the coupling is linear in the oscillator raising and lowering operators, $\hat\kappa\rho = \kappa(\bar n + 1)(a^\dagger a \rho - 2a\rho a^\dagger + \rho a^\dagger a)
+\kappa\bar n(aa^\dagger\rho - 2a^\dagger\rho a + \rho a a^\dagger)$, where $\kappa$ is the oscillator decay rate scaled by $\delta\omega$.

For $\lambda\ll 1$ the scaled rate of inter-SVS switching $W_{\rm sw}/\delta\omega \ll \kappa$. Then over dimensionless time $\sim \kappa^{-1}$ there is formed a quasi-stationary distribution of the oscillator over the states $\psi_n$ localized around the initially occupied SVS. It slowly evolves over time $\sim W_{\rm sw}^{-1}$.

From Eq.~(\ref{eq:master_equation}), if damping is small, so that $\kappa$ is small compared to the  transition frequencies $|g_n-g_{n\pm 1}|/\lambda$, time evolution of the populations $\rho_n$ of quasienergy states is described by the balance equation (\ref{eq:balance_equation}), where
\begin{equation}
\label{eq:transition_rates}
W_{mn}^{\rm (e)}= 2\kappa(\bar n + 1)|a_{nm}|^2,\qquad W_{mn}^{\rm (abs)}= \frac{\bar n}{\bar n+1}W_{nm}^{\rm (e)}.
\end{equation}
Here, $a_{mn} \equiv\langle \psi_m|a|\psi_n\rangle = (2\lambda)^{-1/2}(Q+iP)_{mn}$. 

For $\lambda \ll 1$ one can find $a_{mn}$ using the WKB approximation for $\psi_n(Q)$. A significant simplification comes from the fact that classical trajectories $Q(\tau;g)$ of the system with Hamiltonian $g(Q,P)$ are described by the Jacobi elliptic functions; $Q(\tau;g)$ is double-periodic on the complex-$\tau$ plane, with real period $\tau_p^{(1)}(g)$ and complex period $\tau_p^{(2)}(g)$ \cite{Dykman1988a,Dykman2005}. For $|m-n|\ll \lambda^{-1}$ the matrix element $a_{mn}$ is given by the Fourier $m-n$ component  of the function $a(\tau;g) = (2\lambda)^{-1/2}[Q(\tau;g)+iP(\tau;g)]$\cite{LL_QM81}. This gives
\begin{equation}
\label{eq:rates}
W_{n\,n+k}^{\rm (e)}= \frac{\kappa (\bar n +1)k^2\nu_n^4}{\beta\lambda}
\frac{\exp[ k\nu_n\,{\rm Im}~(2\tau_*-\tau_p^{(2)})]}{|\sinh[ ik\nu_n\,\tau_p^{(2)}/2]|^2}
\end{equation}
Here, $\tau_*\equiv \tau_*(g_n)$ and $\tau_p^{(2)} \equiv \tau_p^{(2)}(g_n)$  [Im~$\tau_*, {\rm Im}~\tau_p^{(2)}>0$]; $\tau_*(g_n)$ is the pole of $Q(\tau;g_n)$  
closest to the real axis;
 $\nu_n =2\pi/\tau_p^{(1)}(g_n)$ is the dimensionless frequency of vibrations in the rotating frame with quasienergy $g_n$. To the leading order in $\lambda$, we have $W^{\rm (e)}_{n\,n+k} = W^{\rm (e)}_{n-k\,n}$. Equation~(\ref{eq:rates}) has to be modified for states very close to the extrema of $g(Q,0)$. 

From Eq.~(\ref{eq:rates}), the ratio of the transition rates $W^{\rm (e)}$ is of the form of a power law, 
\begin{equation}
\label{eq:rates_ratio}
W_{n\,n+k}^{\rm (e)}/W_{n+k\,n}^{\rm (e)}=\xi_n^k, \quad \xi_n=e^{2\nu_n\,{\rm Im}~(2\tau_*-\tau_p^{(2)})}
\end{equation}
 for $|k|\ll \lambda^{-1}$. One can show that $\xi_n<1$. Therefore the rates  of transitions from a state $n$ toward the SVS ($W_{n\,n+k}^{\rm (e)}$ with $k<0$) are larger than away from this state ($k>0$). This is to be expected,  as approaching the SVS corresponds to relaxation. However, even for $T=0$ there are also transitions away from the SVS.

A power-law transition rate ratio corresponds to detailed balance. For $T=0$, for the quasi-stationary distribution $\rho_n^{(0)}$ we have $ \rho_{n+k}^{(0)}/\rho_n^{(0)}=W_{n\,n+k}^{\rm (e)}/W_{n+k\,n}^{\rm (e)}$.

\section{The eikonal approximation}

For arbitrary $T$ we seek the quasi-stationary solution of the balance equation  (\ref{eq:balance_equation}) in the eikonal form 
\begin{equation}
\label{eq:eikonal_form}
\rho_n=\exp(-R_n/\lambda), \qquad R_n\equiv R(g_n).
\end{equation}
%
%
%
Away from the critical temperature region $R(g)$ is smooth and  $R(g_{n+k})-R(g_n)\approx R'(g_n)(g_{n+k}-g_n)$. Then Eq.~(\ref{eq:balance_equation}) is reduced to a polynomial equation for $\exp[-\nu_n|R'(g_n)|]$ \cite{Dykman1988a,Marthaler2006}. This corresponds to the real-time instanton approach in the problem of the distribution of reaction systems \cite{Kamenev2011}. The boundary condition is $R(g_{n=1}) \to 0$ for $\lambda \to 0$; it corresponds to the occupation $\sim 1$ of the closest to the SVS state $n=1$.  Interestingly, $|R'(g_{n=1})|>0$, in contrast to the conventional instanton theory \footnote{V. Peano and M. I. Dykman, in preparation}.

The dimensionless switching rate $W_{\rm sw}$ is determined by the population of states with quasienergies $g_n$ close to the saddle-point quasienergy $g_{\cal S}$ in Fig.~\ref{fig:quasienergy} \cite{Kramers1940}, 
\begin{equation}
\label{eq:W_sw}
W_{\rm sw}\sim \kappa \exp(-R_A/\lambda), \qquad R_A=R(g_{\cal S}).
\end{equation}

For $T=0$ from Eq.~(\ref{eq:rates_ratio}) $|R^{\prime}(g)|\equiv |R^{\prime (0)}(g)|=2{\rm Im}~[\tau_p^{(2)}(g)-2\tau_*(g)]$. This solution is perturbed for $T>0$. For a given state $n$, the perturbation is characterized by the ratio $\varkappa_n(T)$ of the rate of absorption-induced transitions to this state to the rate of transitions from it. For $\bar n\ll 1$ using the unperturbed distribution we have
\begin{equation}
\label{eq:rate_ratio}
\varkappa_n(T)=\frac{\sum\nolimits_mw_{mn}}{\sum\nolimits_m W_{nm}^{\rm (e)}}, \qquad w_{mn}=W_{mn}^{\rm (abs)} \rho_m^{(0)}/\rho_n^{(0)}.
\end{equation}
From Eq.~(\ref{eq:rates}), terms $W_{nm}^{\rm (e)}$ in the denominator in Eq.~(\ref{eq:rate_ratio}) exponentially decay with increasing $|n-m|$. In contrast,  $w_{mn} \propto \exp\left [ 2(n-m) \nu_n f_{}(g_n)\right]$ becomes large  for $n-m \gg 1$ if $f_{}(g_n)>0$, 
\[f_{}(g)  = {\rm Im}~[\tau_p^{(2)}(g)-3\tau_*(g)].\]  
This leads to the divergence of $\ka_n$ for $\lambda\to 0$ and the breakdown of the perturbation theory in $\bar n$. 

We will analyze the distribution about the small-amplitude SVS, see Fig.~\ref{fig:quasienergy}, which is of primary interest for the experiment, cf. \cite{Vijay2009,Mallet2009}.  In this case $f_{}(g)$
monotonically decreases with increasing $g$ and goes through zero for $g=g_{e}$; in our case $g_{e}=1/4$  independent of $\beta$. We will consider the range $\beta < 2/27$, where $g_{\cal S}< g_{e}$. 

Finding $\ka_n$ requires evaluating the rates $W_{n\,m}^{\rm (e)}$, and thus the matrix elements $a_{mn}$, for $n-m\gtrsim \lambda^{-1}$. The overlap integral of the wave functions $\psi_n$ and $\psi_m$ for large $|n-m|$ is exponentially small, which suggests using the WKB approximation \cite{LL_QM81}. The next step is to use the conformal mapping of the complex-$Q$ plane on the complex-$\tau$ plane performed by classical orbits $Q(\tau;g)$. The matrix elements are determined by the behavior of these orbits for $|Q|\to \infty$, which allows calculating both the exponent and the prefactor in $a_{mn}$, and thus gives $w_{mn}$, see Appendix. 

The terms $w_{mn}$ display a maximum as a function of $m$ for $g_m$ closest to $g_e$.  Calculating $\sum_m w_{mn}$ by the steepest descent method, we obtain
\begin{equation}
\label{eq:varkappa_n}
\ka_n=C_n \bar{n}\lambda^{-5/2}\exp\left[2\lambda^{-1}\int_{g_n}^{g_{e}}f_{}(g')
dg'\right];
\end{equation}
$C_n\propto (g_e - g_n)^2$ is independent of $\lambda$ and $T$, see Appendix. 

\section{Correction to the $T=0$ distribution and the breakdown of the instanton approximation}

The correction to the distribution can be sought in the form $R_n=R_n^{(0)}+\Delta_n$. For $|\Delta_n|\ll \lambda$ from Eq.~(\ref{eq:balance_equation})
\begin{equation}
\label{eq:Delta_n_perturbation}
\Delta_n=-\lambda\ka_na_n, \qquad a_n>0, 
\end{equation}
where $a_n\sim 1$. From Eq.~(\ref{eq:varkappa_n}), $|\Delta_n|/\lambda$ increases with $n$ exponentially, $\Delta_n/\Delta_{n-1}\approx\exp[2\nu_nf(g_n)]$. The instanton approximation of smooth $R'(g_n)\approx R^{\prime (0)}(g_n)$ breaks down for $|\Delta_n|/\lambda \sim 1$. Thus the condition $\ka_n \sim 1$, or more precisely, $|\ka_n-1|$ is minimal defines the characteristic breakdown quasienergy $g_n=g_\ka(T)$ for given $T$ and the characteristic breakdown temperature $T_{\ka}\equiv T_{\ka}(g_n)$ for given $g_n$. The thermal perturbation is small for given $g$ provided $T\ll T_{\ka}(g)$.

As $T$ increases it first reaches $T_{\ka}(g)$ for $g=g_{\cal S}$.  Therefore for $T<T_c=T_{\ka}(g_{\cal S})\propto \hbar^2$ the distribution $\rho_n$ and the switching rate $W_{\rm sw}$ are described by the $T=0$-expressions. For $T>T_c$ the $T=0$-approximation applies only for $g-g_{\ka}(T)\gg \lambda$. 

\begin{figure}[th]
\center
\includegraphics[width=76mm]{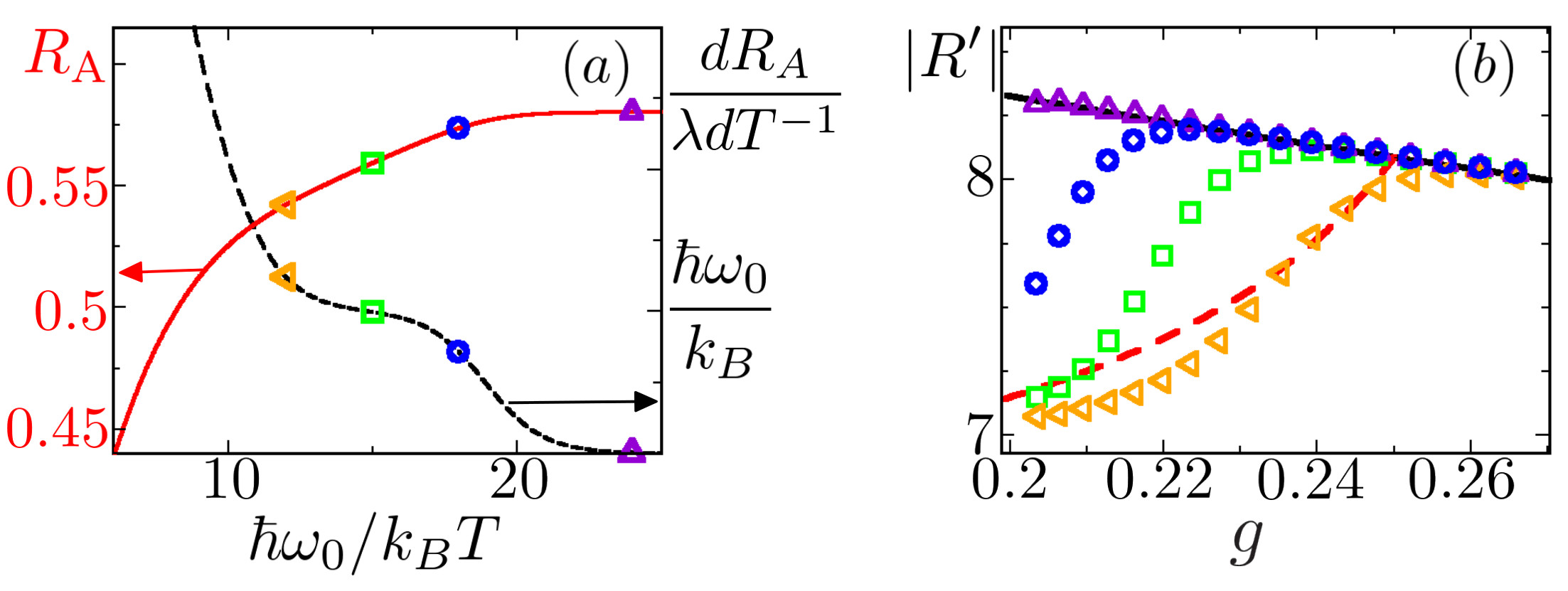}
\caption{Left panel: the activation exponent $R_A$ calculated 
numerically from Eq.~(\ref{eq:balance_equation}) (left curve) and the derivative $dR_A/\lambda dT^{-1}\approx -d\ln W_{\rm st}/dT^{-1}$ (right curve).  The results refer to $\beta=0.0448$ and $\lambda=0.0053$, where the number of localized states is $M=20$. 
Right panel: the steepness $R'=dR/dg$ calculated numerically as $-\nu_n^{-1}\ln(\rho_n/\rho_{n-1})$. The results refer to the temperatures marked on the left panel by the corresponding symbols. The solid and dashed lines show $R'$ for $T=0$ ($R^{\prime (0)}$) and $T\to 0$ ($R'_T$) in the limit $\lambda\to 0$, respectively.}
\label{fig:T_dependence}
\end{figure}

For $g$ between $g_{\ka}(T) $ and $g_{\cal S}$ the distribution is strongly changed by the absorption-induced transitions. From Eqs.~(\ref{eq:balance_equation}) and (\ref{eq:eikonal_form}),  for $g_n,g_m$ below $g_{\ka}$ the corrections to $R^{\prime (0)}(g)$ are no longer small, $\Delta_n-\Delta_m\approx - 2\lambda\nu_nf(g_n)(n-m)$ for $|n-m|\ll 1/\lambda$. The transition from the exponential dependence of $\Delta_n$  on $n$ for $g_n>g_{\ka}$ to this smooth dependence occurs in a narrow range around the value of $n$ where $\ka_n=1$, see Appendix. It corresponds to a kink of $R'(g)$ centered at $g=g_{\ka}(T)$, where $R'(g)$ changes from $\approx R^{\prime (0)}(g)$ to $\approx R'_T(g)= -2{\rm Im}~\tau_*(g)$. Such kink is indeed seen in the numerical data in Fig.~\ref{fig:T_dependence}~(a).  For $\bar n\ll \lambda^3$, absorption-induced transitions to states $n$ with $g_{\ka}-g_n\gg \lambda$ come primarily from states  with $g\approx g_e$, as seen from Eq.~(\ref{eq:varkappa_n}).

With increasing temperature, the kink of $R'(g)$ at $g\approx g_{\ka}$ moves from $g_{\cal S}$ to $g_e$. From Eq.~(\ref{eq:varkappa_n}), it approaches $g_e$ and disappears for $\bar n \sim \lambda^{3/2}$ or $T \sim T_{\rm inst} =\hbar\omega_0/k_B|\ln \lambda|$, still deep in the quantum domain. For higher temperatures but still $\bar n\ll 1$, the slope of $R'(g)$ changes in a narrow region $|g-g_e|\ll 1$ from $R^{\prime (0)}(g_e)$ to $R'_T(g_e)$. A correction to $R'_T(g)$ for $g<g_e$ can be found in a way similar to Ref.~\onlinecite{Marthaler2006}; it is $\sim \bar n^{1/3}$, see Appendix.

From the above arguments, for $T_c<T<T_{\rm inst}$ the effective switching activation energy $R_A\approx \lambda|\ln W_{\rm sw}|$ is
\begin{equation}
\label{eq:activation_energy}
R_A  \approx\int\nolimits_{g_{_{\rm SVS}}}^{g_{\ka}}dg R^{\prime (0)}(g) + \int\nolimits_{g_{\ka}}^{g_{\cal S}}dg R'_T(g),
\end{equation}
($g_{_{\rm SVS}}$ is the quasienergy of the occupied SVS). It depeds on $T$ through $g_{\ka}$. From Eq.~(\ref{eq:varkappa_n}),  $d\ln W_{\rm sw}/dT^{-1} \approx -(\hbar \omega_0/k_B)$ is independent of the modulation. This unexpected behavior is confirmed by numerical simulations, see Fig.~\ref{fig:T_dependence}.

\section{Conclusions}

In conclusion, we have studied the probability distribution and switching of a resonantly modulated nonlinear quantum oscillator. We find a region of quantum crossover, which lies between the critical temperature $T_c\propto \hbar^2$ and $T_{\rm inst}\propto \hbar/|\ln \hbar|$. In this region the slope of the {\it logarithm} of the oscillator distribution over quasienergy states, which is the analog of the effective reciprocal temperature for this distribution, displays a kink. The slope changes from the $T=0$ value, where the dissipation-induced interstate transitions are balanced within a few nearest states, to the value where long-range transitions are important. The kink is not described by the conventional instanton theory, which assumes that the slope of $\ln\rho_n$ is smooth. As a consequence of the kink, in the deeply quantum regime the oscillator switching rate is of the activation form with the activation energy $\hbar\omega_0$ independent of the modulation strength. The results bear on the current experiments on nonlinear cavity modes and modulated Josephson junctions.

We are grateful to G. Sch\"on for an insightful discussion. VP and MID acknowledge support from the NSF, grant EMT/QIS 082985, and the ARO, grant W911NF-12-1-0235.
\appendix

\section{Matrix elements between  remote quasienergy states}

In this Section we calculate in the WKB approximation the exponent and the prefactor of the matrix element $a_{mn}$ of the lowering operator between the eigenstates $\psi_m$ and $\psi_n$ of the quasienergy Hamiltonian $\hat g$ given by Eq.~(4) of the main text. The approach we propose is similar to, but not identical to, that developed \cite{Peano2012} in the problem of the effect of counter-rotating  terms on the dynamics of a parametrically modulated quantum oscillator. 

The Hamiltonian function $g(Q,P)$ has the shape of a tilted Mexican hat, see Fig.~1 of the main text. For the values of $g$ between the local maximum and the saddle point of this surface there are two classical phase-space orbits $P(Q,g)$, which lie on the inner dome and on the external part of $g(Q,P)$. Our analysis refers to the small-amplitude  stable vibrational state SVS, and we count the quantum states off from the state $n=1$, which is closest in quasienergy to the local maximum of $g(Q,P)$.  In the spirit of Ref.~\onlinecite{LL_QM81}
\begin{eqnarray}
\label{eq:matrix_element_defined}
&& a_{mn}=\mathrm{Re}\int_{-\infty}^\infty dQ a_+(Q), \qquad n>m,\nonumber\\
&&a_+(Q)= \psi_m(Q)(Q+ \lambda\partial_Q)\psi^+_n(Q)/\sqrt{2\lambda}\,.
\end{eqnarray} 
Function $\psi^+_n(Q)$ is an eigenfunction of operator $\hat g$ such that $\mathrm{Re}[\psi_n^+(Q)]=\psi_n(Q)$ and that slightly above the real $Q$-axis in the WKB approximation 
\begin{eqnarray}
\label{eq:psi_n+}
 \psi^+_n(Q) \approx  \left(\frac{-2\nu_n}{\pi\partial_P g_n}\right)^{1/2}\exp\left[i\lambda^{-1} S_n(Q) +i\pi/4\right].
\end{eqnarray}
Here, $\partial_Pg_n$ is $\partial_Pg$ calculated for the momentum $P=P(Q,g_n)$; $S_n(Q)$  is the mechanical action counted off from the right turning point $Q_{R}(g_n)$ of the classical orbit that lies on the inner dome of $g(Q,P)$ with $g(Q,P)=g_n$. For the inner-dome orbits the momentum is  
\begin{equation}
\label{eq:momentum}
P(Q,g)=\sqrt{1-Q^2- 2\sqrt{g+\beta^{1/2}Q}},
\end{equation}
and $S_n(Q) =\int_{Q_R(g_n)}^QP(Q',g_n)dQ'$.

The WKB approximation does not apply to function $a_+(Q)$  close to the zeros and the branching points of $P(Q,g_n)$ and $P(Q,g_m)$. We go around these points by lifting  the integration contour above the real $Q$-axis.  

On the real-$Q$ axis the WKB  wave function with quasienergy $g_m$ is $\psi_m(Q)\propto \cos[i\lambda^{-1}S_m(Q)+i\pi/4]$ for $Q$ in the interval between the left and right turning points, $Q_L(g_m)$ and $Q_R(g_m)$. Above this interval on the complex-$Q$ plane one of the terms in the cosine becomes exponentially small and should be disregarded in the WKB approximation \cite{LL_QM81}, so that 
\begin{equation}
\label{eq:psi_m}
 \psi_m(Q) \approx  \left(\frac{-\nu_n}{2\pi\partial_P g_n}\right)^{1/2}\exp\left[-i\lambda^{-1} S_n(Q) -i\pi/4\right].
\end{equation}

We will illustrate the calculation by considering the case $g_m< \sqrt{\beta}$. For such $g$, the orbits on the external part of $g(Q,P)$ have the shape of a horseshoe. Function $P(Q,g_m)$ as given by Eq.~(\ref{eq:momentum}) has a branching point $Q_B=-g/\sqrt{\beta}$, which corresponds to the two utmost negative-$Q$ points of the horseshoe. It also has three zeros: $Q_R(g_m)$, $Q_L(g_m)$, and $Q_{\rm ext}(g_m)$ on the real-$Q$ axis, with $Q_B<Q_{\rm ext}<Q_L<Q_R$. 

\begin{figure}[h]
\includegraphics[width=82mm]{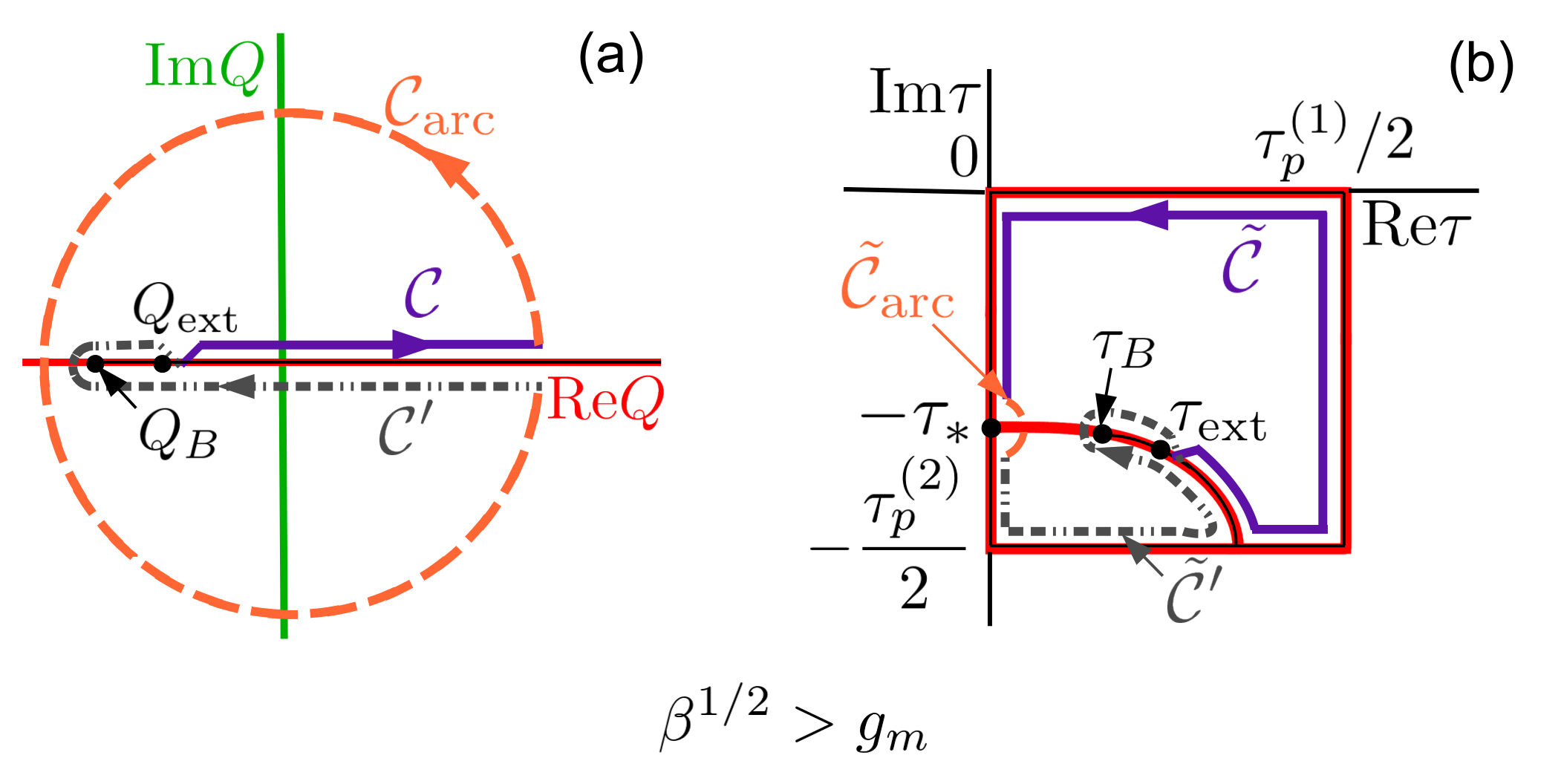}\\
\includegraphics[width=82mm]{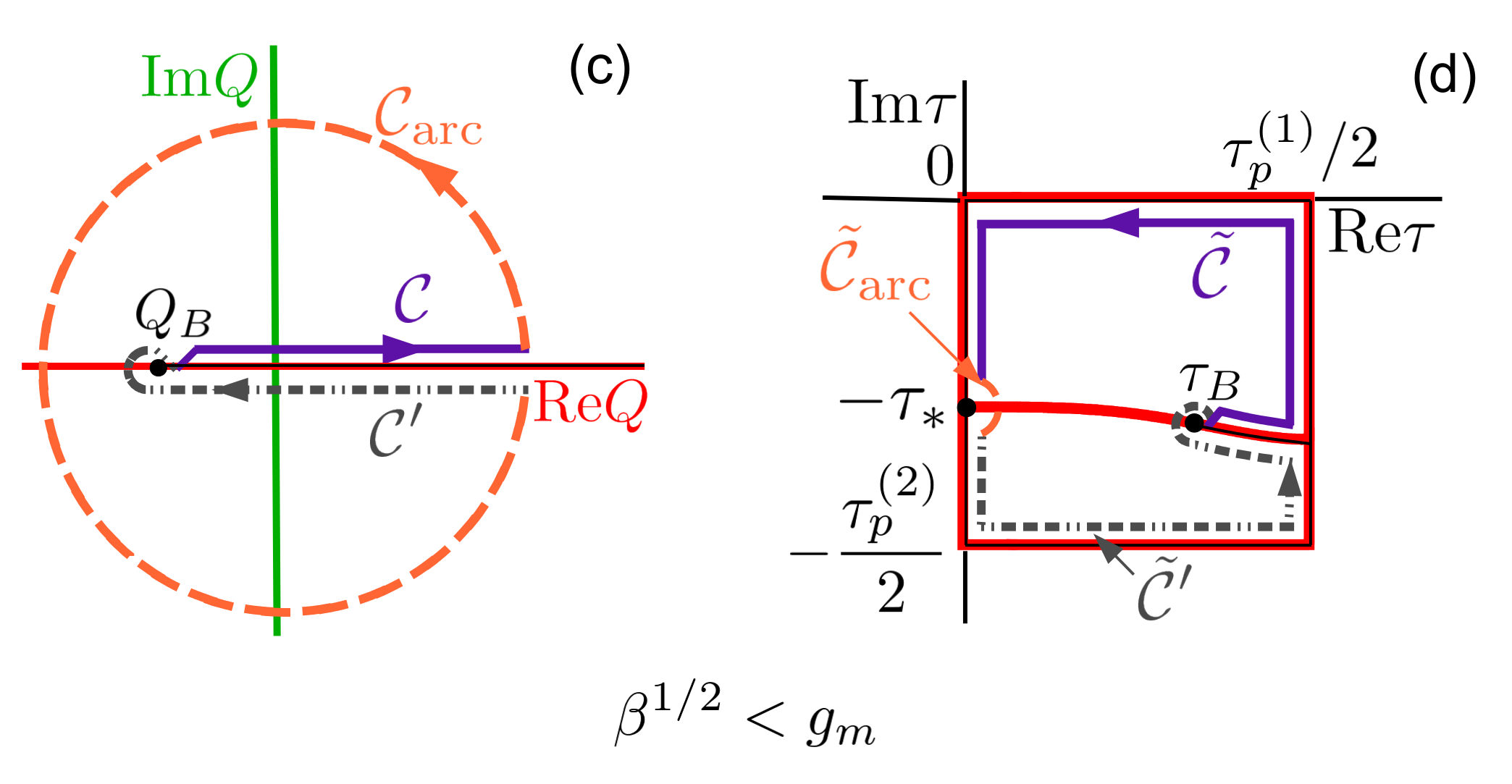}
\caption{(a) The contour of integration ${\cal C}$ for calculating the matrix element (\ref{eq:matelC}) in the WKB approximation and the auxiliary integration contours ${\cal C}'$ and ${\cal C}_{\rm arc}$ for $\sqrt{\beta}>g_m$; $Q_B\equiv Q_B(g_m)$ and $Q_{\rm ext}\equiv Q_{\rm ext}(g_m)$ are the branching point and turning points of $P(Q,g_m)$, see Eq.~(\ref{eq:momentum}). (b) Mapping of the $Q$-plane (with a branch cut from $Q_{\rm ext}$ to $\infty$, the black thin line) on the interior of a rectangle on the $\tau$-plane for $g_m<\sqrt{\beta}$ by function $Q(\tau;g)$ that describes the classical Hamiltonian trajectory with given $g$, $g_n<g<g_m$; $\tau_p^{(1)}$, $\tau_p^{(2)}$, and $\tau_*$ are the real and imaginary periods and the pole of $Q(\tau;g)$, respectively. The solid ($\tilde{\cal C}$), dashed ($\tilde{\cal C}_{\rm arc}$), and dash-dotted ($\tilde{\cal C}'$) lines are the maps of the corresponding contours in (a). The arc in the lower left corner is the map of the real axis of $Q$ from $-\infty$ to $Q_B(g)$; $\tau_B$ and $\tau_{\rm ext}$ are the times for reaching $Q_B(g_m)$ and $Q_{\rm ext}(g_m)$. (c) Integration contours for $g_m>\sqrt{\beta}$. (d) Conformal mapping for $g>\sqrt{\beta}$. The curved red line dividing the rectangle
into two parts  is the map of the real axis of $Q$ from $-\infty$ to $Q_B(g)$}
\label{fig1s} 
\end{figure}

To calculate the integral  in Eq.~(\ref{eq:matrix_element_defined}) we change in the range $Q>Q_{\rm ext}$ to integration over a contour ${\cal C}$ shown in Fig.~\ref{fig1s}~(a). On this contour, from   Eqs.~(\ref{eq:matrix_element_defined}), (\ref{eq:psi_n+}), and  (\ref{eq:psi_m}) $a_+(Q)\approx a^{\rm WKB}_+(Q)$ with
\begin{eqnarray}
\label{eq:h+WKB}
 && a_+^{\rm WKB}(Q) \equiv \pi^{-1}\left(\nu_n\nu_m\bigl/2\lambda\partial_P g_n \partial_P g_m\right)^{1/2}
\nonumber\\
&&\times [Q+ i P(Q,g_n)]\exp\left\{i\left[S_n(Q)-S_m(Q)\right]/\lambda\right\}.
\end{eqnarray}
It is straightforward to show using the full expression for $\psi_m(Q)$ that the real part of the exponent in the expression for $a_+(Q)$ monotonically increases
 on the interval $(-\infty,Q_{\rm ext}]$. Therefore, to logarithmic accuracy the upper bound of the contribution of the integral  from $-\infty$ to $Q_{\rm ext}$ of $a_+(Q)$ in Eq.~(\ref{eq:matrix_element_defined}) is $\sim a_+(Q_{\rm ext})\sim a^{\rm WKB}_+(Q_{\rm ext})$. Below we show that $a_{mn}$ is exponentially larger than $a_+^{\rm WKB}(Q_{\rm ext})$. Then
\begin{eqnarray}
\label{eq:matelC}
a_{mn}\approx\mathrm{Re}\int_{\cal C} dQ a_+^{\rm WKB}(Q).
\end{eqnarray} 

To evaluate the integral  (\ref{eq:matelC}), we analytically continue $a_+^{\rm WKB}$ to the $Q$-plane with a branch cut on the semi-infinite interval  $(Q_B(g_m),\infty)$. We then can change from integration along ${\cal C}$ to integration along the  circle ${\cal C}_{\rm arc}$  and contour ${\cal C'}$ shown Fig.~\ref{fig1s}~(a). 

Classical trajectories $Q(\tau;g)$  for the Hamiltonian $g(Q,P)$ are expressed in terms of  the Jacobi elliptic functions \cite{Dykman1988a}. For each $g$, this expression provides conformal mapping of the $Q$-plane (with a branch cut)  onto a $g$-dependent region on the plane of complex time $\tau$. We define $\tau(Q,g)$ as the duration of classical motion from the turning point $Q_R(g)$ to $Q$. Then  the region of the $\tau$-plane that corresponds to the $Q$-plane (with a branch cut) is the interior of a rectangle shown in Fig.~\ref{fig1s}~(b).

Since $\tau(Q,g)=\partial S/\partial g$, the exponent in Eq.~(\ref{eq:h+WKB}) is
\begin{equation}
\label{eq:sn-s0}
\frac{i}{\lambda}[ S_n(Q)-S_m(Q)]=-\frac{i}{\lambda}\int_{g_n}^{g_m}\!\!dg\tau(Q,g)\,.
\end{equation}
As seen in Fig.~\ref{fig1s}~(b), for any $g$ between $g_n$ and $g_m$,  for any $Q_{\rm arc}$ on contour   ${\cal C}_{\rm arc}$ and any $Q'$ on contour ${\cal C'}$,  $|$Im~$\tau(Q_{\rm arc},g)| <$~$|$Im~$\tau(Q',g)|$. Therefore, $a_+^{\rm WKB}(Q)$ is exponentially smaller on contour ${\cal C}'$  than on contour ${\cal C}_{\rm arc}$, and the integral along  ${\cal C'}$ can be disregarded. Moreover, $a_+^{\rm WKB}(Q_{\rm ext}(g_m))\ll a_{mn}$, as assumed in Eq. (\ref{eq:matelC}).

The integral along ${\cal C}_{\rm arc}$  can be evaluated using the asymptotic expressions $P\approx -iQ$,
$g_P\approx i2\beta^{1/4}Q^{3/2}$,
\begin{equation}
\label{eq:sn-s02}
\frac{i}{\lambda}[ S_n(Q)-S_m(Q)]\approx\frac{i}{\lambda}\int_{g_n}^{g_m}\!\!dg\tau_*(g)+\frac{g_m-g_n}{\lambda\beta^{1/4}Q^{1/2}}.\nonumber
\end{equation} 
With Eq.~(\ref{eq:sn-s02}) the integral (\ref{eq:matelC}) is reduced to a simple residue, which gives
\begin{equation}\label{eq:matelan}
 a_{mn}=\left(\frac{2\nu_n\nu_m}{\beta\lambda^3 } \right)^{1/2}
(g_m-g_n)\exp\left[i\lambda^{-1}\int_{g_n}^{g_m}\!\!dg\tau_*(g)
\right]
\end{equation}

For $g_m> \sqrt{\beta}$,  the small-momentum branch $P(Q,g_m)$ (\ref{eq:momentum}) has only two turning points.  In the WKB approximation,  $a_+(Q)\approx a_+^{\rm WKB}(Q)$ on contour ${\cal C}$ whose left endpoint is the branching point $Q_B(g_m)$, see Fig.~\ref{fig1s}~(c). The conformal mapping $Q(\tau;g)$ has a different topology, which is  shown in Fig.~\ref{fig1s}~(d). Nonetheless, using the same arguments as before, we arrive at the same expression (\ref{eq:matelan}) for the matrix elements $a_{mn}$.

\begin{figure}[h]
\hspace*{1cm}
\begin{center}
\includegraphics[height=35mm]{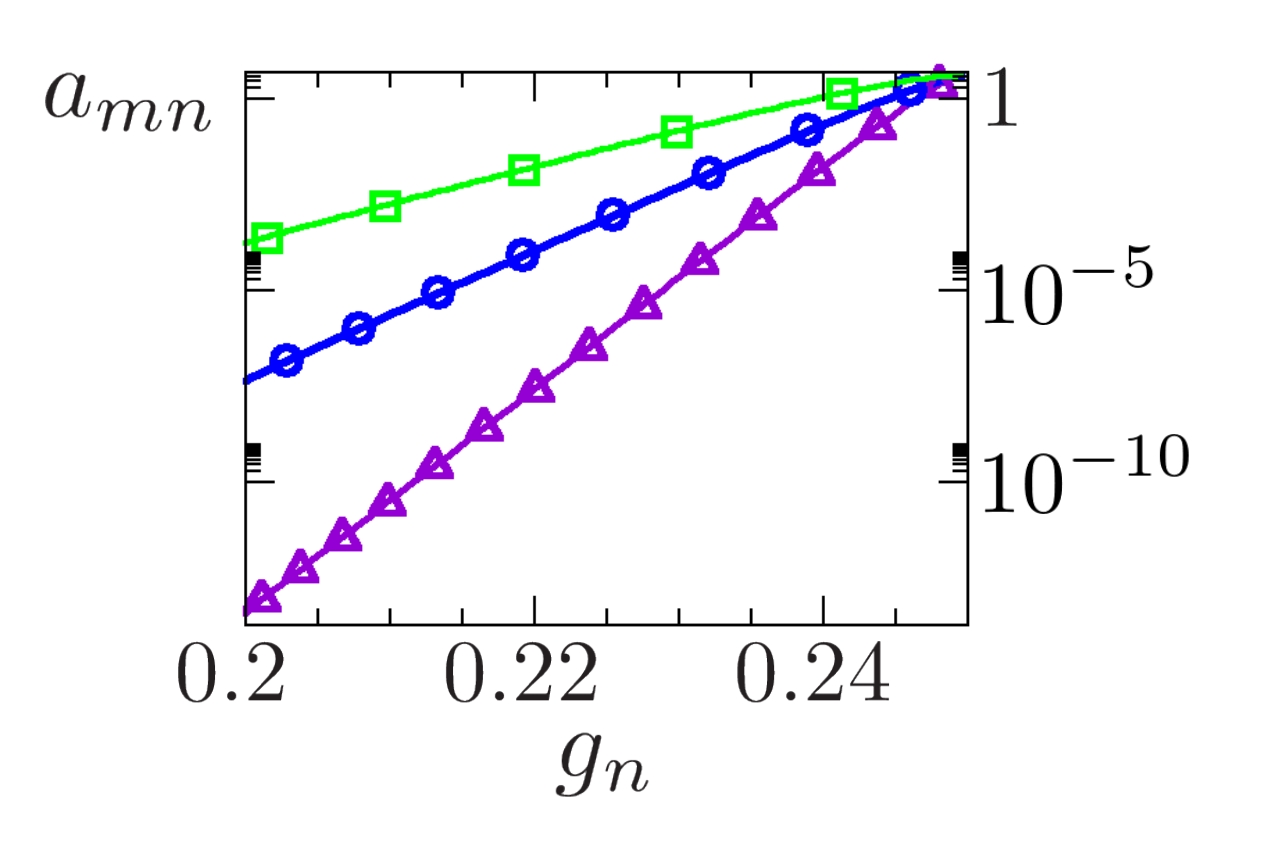}
\caption{Comparison of Eq.\ (\ref{eq:matelan}) for  $a_{mn}$ calculated as a continuous function of $g_n$ (solid lines) with  numerical calculations (symbols). The scaled intensity of the modulating field is  $\beta=0.0448$. The other parameter values are $\lambda=0.0053$ and $m=4$  (violet triangles), $\lambda=0.09$ and $m=2$ (blue circles), and  $\lambda=0.15$ and $m=1$ (green squares). Parameter $m$ has been chosen so that $g_m\approx g_{e}\equiv 1/4$.}
\label{fig2s} 
\end{center}
\end{figure}

In Fig.~\ref{fig2s} we compare the explicit analytical expression for the matrix elements $a_{mn}$, which includes both the exponent and the prefactor, with the numerical calculations based on solving the Schr\"odinger equation $\hat g\psi_n = g_n\psi_n$. The results are in excellent agreement.

\section{Thermally-induced modification of the $T=0$ distribution}

The explicit expression for the transition matrix elements (\ref{eq:matelan}) makes it possible to calculate the rate of absorption-induced transitions to a state $n$ scaled by the rate of leaving this state  $\ka_n(T)$, Eq.~(11) of the main text. The most probable transitions are those from states $m$ closer to the SVS, $m<n$. From Eq.~(\ref{eq:matelan}), the term $w_{mn}=W_{mn}^{\rm (abs)}\rho_m^{(0)}/\rho_n^{(0)}$ in the expression for $\ka_n$ for  $n-m\gg 1$ is of the form
\begin{eqnarray}
\label{eq:w_mn}
&& w_{mn} \approx \bar n \kappa\lambda^{-3}C^{(w)}(g_m,g_n)\exp\left[2\lambda^{-1}\int_{g_{n}}^{g_{m}}f(g)dg\right],\nonumber\\
&&C^{(w)}(g_m,g_n) = (4\nu_m\nu_n/\beta)\,(g_m-g_{n})^2 .
\end{eqnarray}
As a function of $m$, $w_{mn}$ is maximal for $g_m$ closest to the quasienergy value $g_e$ given by the condition $f(g_e)=0$.

The leading-order contribution to $\ka_n$ comes from the terms $w_{mn}$ with $|g_m - g_e|\ll 1$. The sum of these terms can be calculated by changing to integration over $m$ and then using the steepest descent method. This gives Eq.~(12) of the main text for $\ka_n$, with 
\begin{eqnarray}
\label{eq:C_n_coefficient}
C_n &=& C^{(w)}(g_e,g_n)[\kappa^2\pi/f'(g_e)]^{1/2}/\left(\lambda W_n^{(e)}\right),\nonumber\\
&&W_n^{\rm (e)}=\sum\nolimits_mW_{nm}^{\rm (e)}.
\end{eqnarray}
Here, $f' = df/dg$, whereas $W_n^{\rm (e)}$ is the overall rate of transitions from state $n$ for $\bar n \ll 1$. It can be found from Eq.~(7) of the main text; using this equation we obtain $C_n\sim C^{(w)}(g_e,g_n)[\pi/f'(g_e)]^{1/2}$.

We now consider the distribution for $T>0$, 
\begin{equation}
\label{eq:rho_with_modification}
\rho_n=\exp(-R_n/\lambda), \qquad R_n=R_n^{(0)} + \Delta_n.
\end{equation}
 For small $\bar n$  we need to consider the absorption-induced transitions to a given state $n$ only from states $m$ with $m<n$. Then the balance equation in the quasi-stationary regime reads
\begin{eqnarray}
\label{eq:balance_with_correction}
&&\sum\nolimits_m\left(W_{nm}^{\rm (e)}/W_n^{\rm (e)}\right)e^{(\Delta_n-\Delta_m)/\lambda} - 1= -\wp_n, \nonumber\\
&&\wp_n= \sum\nolimits_{m<n} \left(w_{mn}/W_n^{\rm (e)}\right)e^{(\Delta_n-\Delta_m)/\lambda}.
\end{eqnarray}
The parameter $\ka_n$ is given by $\wp_n$ for $\Delta_n=0$.

\subsection{Temperature range $\bar n \ll \lambda^3$}

Thermal modification of the distribution becomes substantial when $\bar n$ is exponentially small in $\lambda$, $|\ln \bar n|\gtrsim R^{(0)}(g_{\cal S})/\lambda$. In this subsection we consider the distribution for  $\bar n \ll \lambda^3$. 

We start with the range  of the level numbers $n$ where $n_{\ka}-n\gg 1$; the quasienergy level number $n_{\ka}$ is given by the condition that $|\ka_{n_{\ka}}-1|$ be minimal, $g_{n_{\ka}} \equiv g_{\ka}(T)$. For such $n$, $|\Delta_n|/\lambda \ll 1$, and in $\wp_n$ one can disregard $\Delta_n, \Delta_m$. Then $\wp_n = \ka_n$ is given by Eq.~(12) of the main text.  In the left-hand side of Eq.~(\ref{eq:balance_with_correction}) it suffices to keep the terms linear in $\Delta_n, \Delta_m$.  Taking into account that $\ka_n/\ka_{n-1}  \approx \exp[2\nu_nf(g_n)]$, from Eq.~(\ref{eq:balance_with_correction}) we then obtain Eq.~(13) of the main text with 
\[a_n \approx \left\{\sum\nolimits_m\left( W_{nm}^{\rm (e)}/W_n^{\rm (e)}\right)\left[ 1 - e^{2\nu_n (m-n) f(g_n)}\right]\right\}^{-1}.
\]
Using the explicit form of $W_{nm}^{\rm (e)}$ and taking into account that Im~$(\tau_p^{(2)} - 2\tau_*) > |{\rm Im}~(\tau_p^{(2)} - 4\tau_*)|$, one can show  that $a_n >0$; clearly, $a_n$ is independent of $\lambda$ and $\bar n$.

We then consider the states with $n>n_{\ka}$, and first assume that $n-n_{\ka}\ll 1/\lambda$. In this range it is convenient to split the sum over $m$ in $\wp_n$ in Eq.~(\ref{eq:balance_with_correction}) into a sum from $n-1$ to $n_{\ka}-n_0$ and a sum over $m<n_{\ka}-n_0$, where $n_0$ is chosen so that $1 \ll n_0 \ll 1/\lambda$. Since $\Delta_m$ is exponentially small for $n_{\ka}-m\gg 1$, it can be dropped in the terms with $m<n_{\ka}-n_0$. Taking into account the explicit form of $w_{mn}$, Eq.~(\ref{eq:w_mn}), and that, as a consequence of this equation,  $\ka_n$ depends on $n$ exponentially, we write $\wp_n$ as 
\begin{eqnarray*}
\label{eq:split_sum}
&&\wp_n=\wp_n^{(1)}+\wp_n^{(2)}, \\
&&\wp_n^{(1)}=\sum\nolimits_{m=n_{\ka}-n_0}^{n-1} \left(w_{mn}/W_n^{\rm (e)}\right) e^{(\Delta_n-\Delta_m)/\lambda},\\
&&\wp_n^{(2)} \approx e^{(\Delta_n/\lambda) +2(n-n_{\ka})\nu_{n}f(g_{n})}\ka_{n_{\ka}}.
\end{eqnarray*}
From the relation  $\ka_{n_{\ka}}\sim 1$, it follows that the solution of Eq.~(\ref{eq:balance_with_correction}) is 
\begin{eqnarray}
\label{eq:solution_Delta_n}
&&\Delta_n-\Delta_{m}\approx -2\lambda\nu_n (n-m) f(g_n), n-m\ll \lambda^{-1},\nonumber\\
&&\Delta_n\approx - 2\lambda (n-n_{\ka})f(g_n)
\end{eqnarray} 
provided $n-n_{\ka}>m-n_{\ka} \gg 1$. Corrections to $\Delta_n$ come from the terms $\Delta_{n'}$ with $n' -n_{\ka} \sim 1$, $|\Delta_{n'}|\sim \lambda$. They also come from the prefactor in $w_{mn}$,  the dependence of $\nu_n$ and $ g_n$ on $n$, and the dependence on $n$ of the left-hand side of Eq.~(\ref{eq:balance_with_correction}) calculated for $\Delta_n-\Delta_m$ of the form (\ref{eq:solution_Delta_n}). All these correction are $\sim\lambda$. For $\Delta_n$ of the form (\ref{eq:solution_Delta_n}) $\wp_n^{(1)} \lesssim \bar n/\lambda^3  \ll \wp_n^{(2)}$. 

In contrast to the exponentially steep $n$-dependence  of $\Delta_n=-\lambda a_n\ka_n$ for $n<n_{\ka}$, in Eq.~(\ref{eq:solution_Delta_n}) $\Delta_n$ smoothly depends on $n$. The region of $n$ where the two expressions join one another is centered at $n_{\ka}$. The width of this region is independent of $\lambda$ or $\bar n$, as seen from Eq.~(\ref{eq:balance_with_correction}). In this region $\Delta_n/\lambda \sim 1$.

One can easily see that Eq.~(\ref{eq:solution_Delta_n}) applies also for $n-n_{\ka}\sim 1/\lambda$. Here, too, $\wp_n^{(1)}\lesssim \bar n (n-n_{\ka})^3 \lesssim \bar n/\lambda^3 \ll \wp_n^{(2)}$; in the expression for $\Delta_n$ one should replace
\[2(n-n_{\ka})\nu_{n}f(g_{n}) \to 2\lambda^{-1}\int_{g_n}^{g_{\ka}}dg f(g).\]
The inequality $\wp_n^{(1)}\ll \wp_n^{(2)}$ indicates that in the temperature range $\bar n\ll \lambda^3$ absorption-induced transitions to states $n>n_{\ka}$ come primarily from remote states with quasienergy $\approx g_e$. 

\subsection{Temperature range $\bar n \gg \lambda^3$}

As the temperature increases $\bar n$ becomes larger than $\lambda^3$
and the  term $\wp_n^{(1)}$ becomes more important. Equation~(\ref{eq:solution_Delta_n}) still gives the leading-order term in $(\Delta_n-\Delta_m)/\lambda$ in the range $n-n_{\ka}, m-n_{\ka} \gg \bar n^{-1/3}$. However,  when calculating $\wp_n^{(1)}$ one should add a correction $-(n-m)\epsilon_n$  to this term, $|\epsilon_n|\ll 1$. From Eq.~(\ref{eq:w_mn}), the sum over $m$ in $\wp_n^{(1)}$ is then a second derivative with respect to $\epsilon_n$ of a geometric series, since in $C^{(w)}$ we have $(g_n-g_m)^2\approx [\lambda\nu_n (n-m)]^2$. The summation gives 
\begin{eqnarray}
\label{eq:wp2}
\wp_n^{(1)}\approx (\bar n/\epsilon_n^3)C_{\wp n}^{(1)},\qquad C_{\wp n}^{(1)}=8\nu_n^4\kappa/\lambda\beta W_n^{\rm (e)}.
\end{eqnarray} 
The coefficient $C_{\wp n}^{(1)}$ is independent of $\lambda$ and is $\sim 1$, generally. Since we need $\wp_n <1$,  from Eq.~(\ref{eq:wp2}) we have $\epsilon_n\sim \bar n^{-1/3}$. Formally, the sum over $n-m$ in $\wp_n^{(1)}$ was extended to infinity, which requires that $n-n_{\ka}\gg 1/\epsilon_n$, and then from Eq.~(\ref{eq:solution_Delta_n}) $\epsilon_n \gg \lambda$.

The correction to $\Delta_n/\lambda$ in the expression for $\wp_n^{(2)}$ is given by $\sum_{k< n}\epsilon_n$. It becomes $\sim -\bar n^{-1/3}/\lambda\gg 1$ for $n-n_{\ka}\sim \lambda^{-1}$. Therefore $\wp_n^{(2)}$ is small for such $n$. On the whole, as the ratio $\bar n^{1/3}/\lambda$ changes from small to large, so does also the ratio $\wp_n^{(1)}/\wp_n^{(2)}$, starting first from large $n$ and then going to smaller and smaller $n$. 

From Eq.~(\ref{eq:w_mn}) and from Eq.~(12) of the main text, for $\bar n$ approaching $\lambda^{3/2}\ll 1$ the quasienergy $g_{\ka}$ approaches $g_e$. For $\bar n\gg \lambda^{3/2}$ ($T>T_{\rm inst}$) the rates of transitions into  all states with $g_n<g_e$ are determined by thermal processes. In this range  $\wp_n=\wp_n^{(1)}$  and the modification of $R_n$ compared to $R_n^{(0)}$ is determined by Eq.~(\ref{eq:solution_Delta_n}). 

Keeping in mind that $R_n-R_{n-1}\approx  -\lambda R'(g_n)\nu_n$ away from $n\approx n_{\ka}$, we obtain from Eqs.~(\ref{eq:rho_with_modification}) and (\ref{eq:solution_Delta_n}) 
\begin{equation}
\label{eq:thermal_R'}
R'(g)\approx R'_T(g)= -2{\rm Im}~\tau_*(g), \qquad g<g_{\ka}. 
\end{equation}
Thus in the whole transition region where $g_{\cal S}< g_{\ka} < g_e$ (and respectively, $T_c<T < T_{\rm inst}$) the distribution depends  on temperature primarily through the position $g_{\ka}$ of the kink of $R'(g)$, where $R'$ sharply changes from $R^{\prime (0)}$ to $R'_T$.


%

\end{document}